\documentclass[%
 reprint,
groupedaddress,
 amsmath,amssymb,
 aip,
apl,
]{revtex4-1}

\usepackage{graphicx}
\usepackage{dcolumn}
\usepackage{bm}
\usepackage{hyperref}
\usepackage[mathlines]{lineno}
\usepackage{siunitx}
\usepackage{xspace}
\usepackage{epstopdf}
\usepackage{titlesec}
\titlespacing\subsubsection{0pt}{12pt plus 4pt minus 2pt}{6pt plus 2pt minus 2pt}
\usepackage{multirow}

\usepackage[caption=false]{subfig}

\newcommand{\NGO}{$\mathrm{NdGaO_3}$\xspace}
\newcommand{\YBCO}{$\mathrm{YBa_2Cu_3O_{7-\delta} }$\xspace}

\newcommand{\tc}{$T_\textrm{c}$}
\newcommand{\LSCO}{$\mathrm{La_{2-x}Sr_xCuO_4}$\xspace}

\begin{document}

\title{Coherently strained epitaxial YBa$_2$Cu$_3$O$_{7-\delta}$ films grown on NdGaO$_3$ (110)}

\author{Sogol Khanof}
\author{Jochen Mannhart}
\author{Hans Boschker}%
\email{h.boschker@fkf.mpg.de}
\affiliation{%
Max-Planck-Insitute for Solid State Research,\\
70569 Stuttgart, Germany
}%

%
%

\date{\today}

\begin{abstract}
\YBCO is a good candidate to systematically study high-temperature superconductivity by nanoengineering using advanced epitaxy. An essential prerequisite for these studies are coherently strained \YBCO thin films, which we present here using \NGO (110) as a substrate. The films are coherent up to at least 100 nm thickness and have a critical temperature of 89$\pm$1 K. The $a$ and $b$ lattice parameters of the \YBCO are matched to the in-plane lattice parameters of \NGO (110), resulting in a large reduction of the orthorhombicity of the \YBCO. These results imply that a large amount of structural disorder in the chain layers of \YBCO is not detrimental to superconductivity.

\end{abstract}

\maketitle

The origin of high-temperature superconductivity in the cuprates is one of the biggest and most fascinating challenges in solid-state research. Thin film epitaxy of cuprates superconductors is essential for applications such as superconducting cables \cite{Melhem2012} and devices\cite{Khare2003}. Epitaxial growth is also a vital tool to answer fundamental questions about the nature of superconductivity. For example, the first phase-sensitive determination of the d-wave symmetry of the superconducting order parameter was performed by analyzing the spontaneous flux of \YBCO rings grown on a tricrystal substrate\cite{Tsuei1994}, the amount of admixtures to the d-wave order parameter symmetry were precisely determined using thin film \YBCO-Nb junction technology\cite{Smilde2005, Kirtley2006}, and a direct correspondence between the critical temperature (\tc) and the superfluid density was recently observed in overdoped \LSCO using high-quality epitaxial thin films \cite{Bozovic2016}. 

Moreover, advances in thin film epitaxy such as composition control and layer-by-layer growth monitoring allow for the systematic modification of materials and for artificial materials design\cite{Bousquet2008, Hwang2012, Boschker2012, Lee2013}. \YBCO is a well-studied cuprate superconductor because of its high critical temperature, the ability to grow high-quality single crystals with effective doping control, and its layered crystal structure where dopants reside away from the CuO$_2$ plane, thus minimizing electron scattering\cite{Ghiringhelli2012, Mankowsky2014, Keimer2015c, Badoux2016}. Therefore, \YBCO is an excellent candidate material for studies by epitaxial modification schemes. An essential prerequisite for such studies is a suitable substrate that results in coherently strained \YBCO films that are single domain together with bulk-like superconducting properties. Here, we demonstrate that \YBCO films grown on \NGO (110) fulfill these requirements. 
 
\begin{figure}
\includegraphics[width=0.45\textwidth]{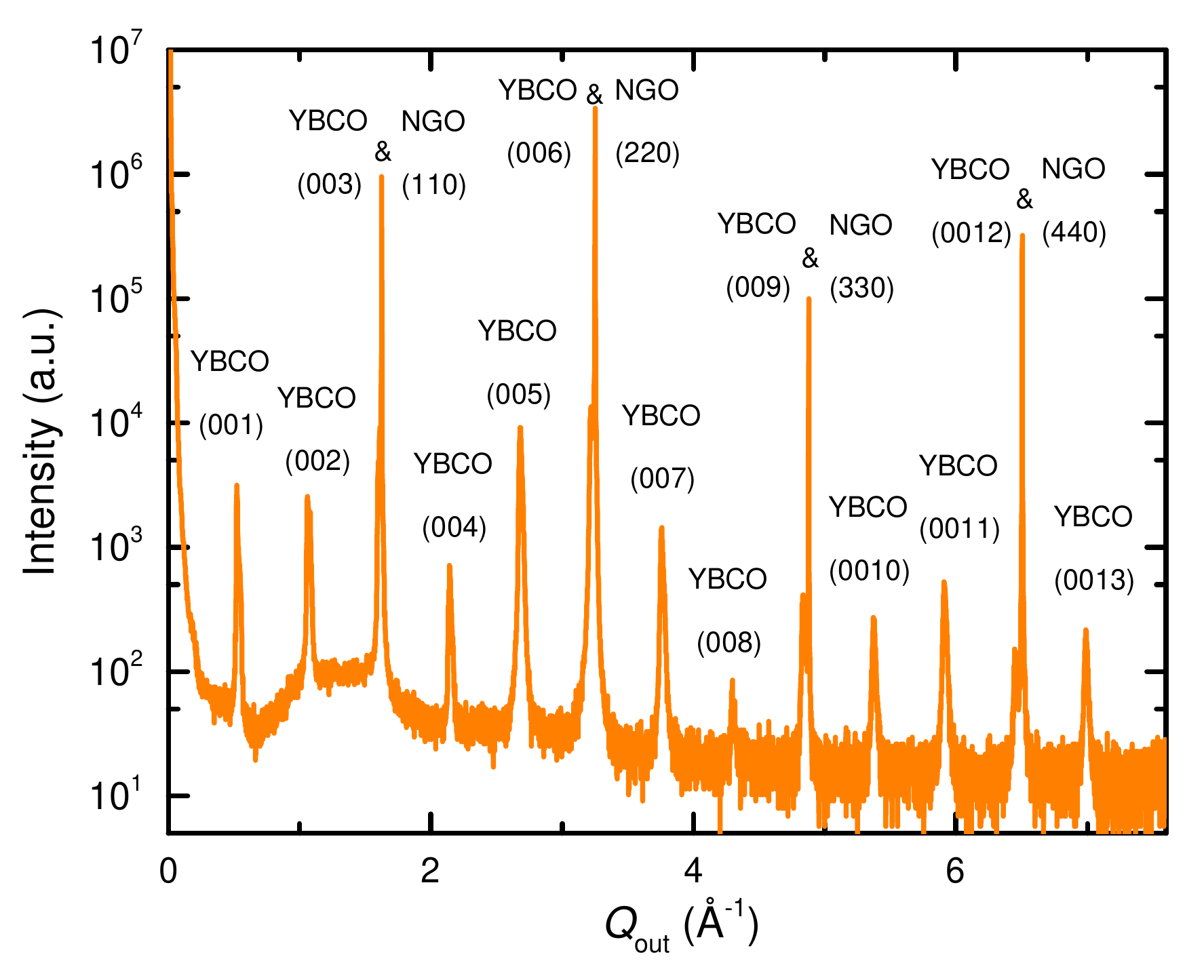}
\caption[]{Out-of-plane momentum scan of the 100 nm-thick \YBCO film. The substrate and film peaks are indicated. Only (00$l$) \YBCO reflections are found, indicating single-phase \YBCO with a $c$-axis orientation. }
\label{t2t}
\end{figure}

\begin{figure*}
\includegraphics[width=0.95\textwidth]{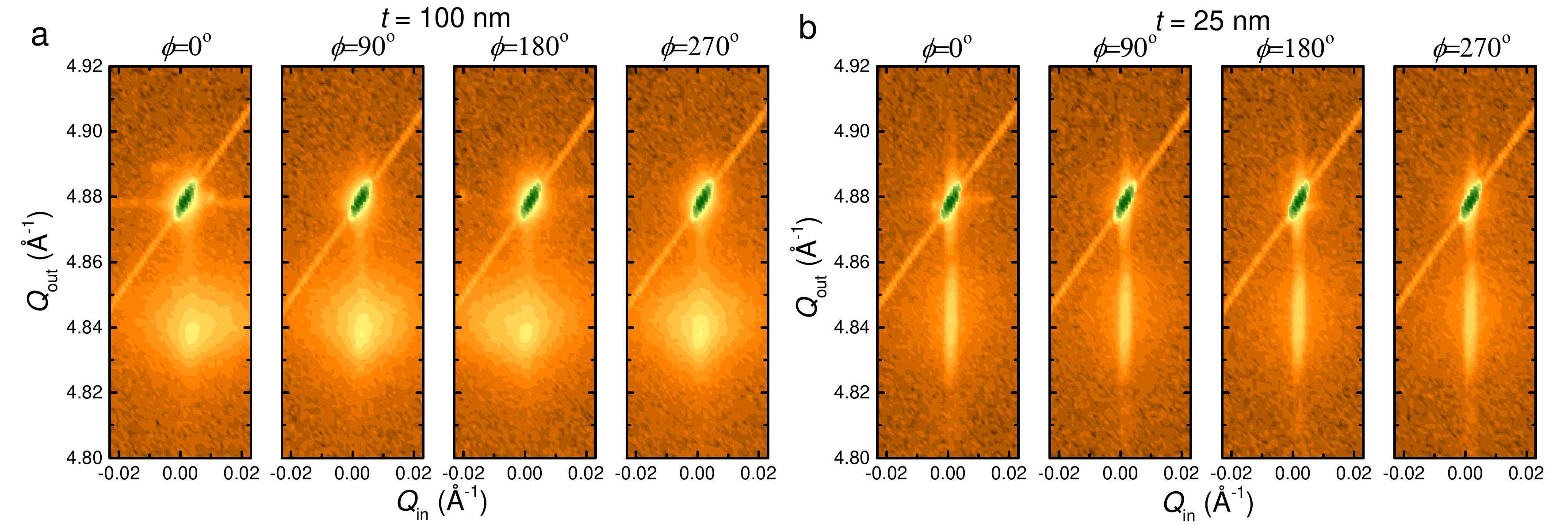}
\vspace{0.5cm}
\includegraphics[width=0.95\textwidth]{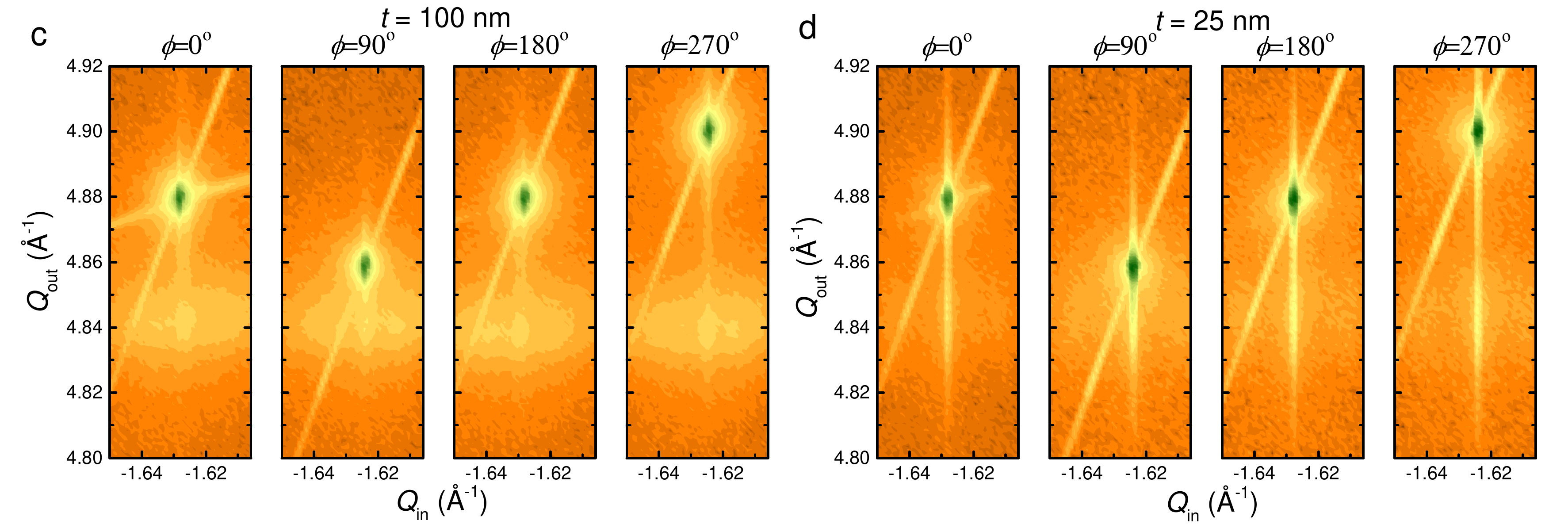}
\caption[]{Reciprocal space maps for different values of the angle $\phi$ around (a,b): the (009) reflection, and (c,d): the \{109\} reflections. The 100-nm-thick film has a significant mosaic spread and is not perfectly aligned with the \NGO crystal. In contrast, the 25-nm-thick film is fully coherent with the underlying \NGO. }
\label{rsm009}
\end{figure*}

\NGO has the orthorhombic lattice structure with $a$ = 5.417 \AA, $b$ = 5.499 \AA, and $c$ = 7.717 \AA. Its (110) surface has a rectangular lattice structure with a pseudocubic lattice parameter of 3.85 \AA~along the (001) direction and a pseudocubic lattice parameter of 3.86 \AA~along the (1$\overline{1}$0) direction. Atomically smooth surfaces were obtained by a combination of sonification in deionized water, chemical etching with a buffered hydrofluoric acid solution and a two-hour 1000~$^\circ$C annealing step, similar to the procedure commonly used for SrTiO$_3$ \cite{Koster1998}. The \YBCO samples were grown by pulsed laser deposition from a stoichiometric target. The growth temperature was 780~$^\circ$C, the oxygen partial pressure was kept to 0.4 mbar, the target-substrate distance was 5 cm, and the laser fluence was 3 J/cm$^2$. After growth, the samples were annealed for two hours at 450~$^\circ$C in an oxygen atmosphere of 400 mbar. Four samples were grown with thicknesses $t$ of 10, 25, 50, and 100 nm. The surface morphology was measured with a Cypher atomic force microscope (AFM) and the x-ray diffraction measurements (XRD) were performed with a Panalytical Empyrean diffractometer in high-resolution mode. We define the out-of-plane component of the momentum transfer $Q_\textrm{out}$ as the component of $Q$ parallel to the (110) direction of the \NGO crystal. The in-plane component of the momentum transfer $Q_\textrm{in}$ is the component of $Q$ perpendicular to $Q_\textrm{out}$ and within the diffraction plane. The film thicknesses were determined using a Dektak surface profiler after etching part of the samples with a phosphoric acid solution. The temperature dependence of the resistance was measured in the van der Pauw geometry using a Quantum Design physical properties measurement system. 

Figure~\ref{t2t} shows an out-of-plane XRD scan of the 100 nm-thick sample. Next to four substrate peaks, thirteen peaks corresponding to the (00$l$) reflections of \YBCO are observed. We do not observe peaks corresponding to other orientations of \YBCO, nor those of secondary phases. Therefore we conclude that the \YBCO is single-phase, (001)-oriented. The $c$-axis lattice parameter is 11.7 \AA, in good agreement with the $c$-axis lattice parameter of optimally doped \YBCO single crystals. The precise orientation of the films was measured by performing reciprocal space maps around the (009) reflection for four different values of the azimuthal angle $\phi$, as shown in Fig.~\ref{rsm009}a. The maps also show the \NGO (330) reflection. The film peak is significantly broadened along $Q_\textrm{in}$ in comparison to the substrate peaks for all values of $\phi$. For $\phi$=0$^\circ$ ($\phi$=180$^\circ$), peak center is observed at a larger (smaller) $Q_\textrm{in}$ value than that of the substrate peak and the peak shape is asymmetric with a long tail towards larger (smaller) $Q_\textrm{in}$ values. In contrast, for $\phi$=90$^\circ$ and $\phi$=270$^\circ$, the film peaks are symmetrical and at the same value of $Q_\textrm{in}$ as the substrate peak. This implies the CuO$_2$ planes of the \YBCO are oriented at an angle of 0.07 degrees with respect to the crystal lattice of the \NGO substrate. The in-plane projection of the tilt is mostly along $\phi$ = 0. This tilt exactly corresponds to the miscut of the substrate. Therefore the film is oriented with its [001] lattice direction parallel to the surface normal vector of the \NGO crystal. The alignment of the \YBCO lattice planes with the optical surface of the substrate was also observed for the 50 nm-thick sample. The large peak width along $Q_\textrm{in}$ can be due to either mosaicity or a reduction of the in-plane coherence length. The two scenarios can be discriminated by measuring the peak widths at a series of (00$l$) peaks. This analysis (data not shown) clearly determined the width to be due to a mosaicity of 0.12$^\circ$ with negligible contribution of a reduction of the in-plane coherence length. In contrast to the thick samples, the two thinner films are oriented such that the \YBCO lattice planes are aligned to the \NGO crystal planes, with a peak width smaller than the resolution of the diffractometer, Fig.~\ref{rsm009}b.  

We next turn to the analysis of the in-plane components of the crystal structure. $\phi$-Scans of asymmetric reflections confirmed the epitaxial alignment of all the films with the $a$- and $b$-axes of the \YBCO aligned with the [001] and [1$\overline{1}$0] lattice directions of the \NGO, respectively. Furthermore, we performed reciprocal lattice maps around asymmetric reflections to study the in-plane crystal structure of the films. Representative maps around the set of \YBCO \{109\} reflections are shown in Figs.~\ref{rsm009}c,d. Figure~\ref{rsm009}c shows the maps of the $t$ = 100 nm sample with the in-plane momentum transfer aligned with the \NGO [001], [$\overline{1}1$0], [00$\overline{1}$], and [1$\overline{1}$0] lattice directions, respectively. Broad film peaks are observed, indicating a reduced in-plane coherence of the crystal structure compared to that of the substrate, in agreement with the mosaicity described above. The in-plane lattice parameters were extracted from the differences in peak positions with positive and negative in-plane momentum transfer. We find that the film's in-plane lattice parameters are equal to those of the substrate, $a$ = 3.85 \AA~and $b$ = 3.86 \AA. Figure~\ref{rsm009}d shows the maps of the $t$ = 25 nm sample. For this sample peaks are observed that are very narrow along the in-plane direction and elongated along the out-of-plane direction. The in-plane momentum transfer of the film peaks is equal to that of the substrate peaks. This agrees well with a thin film that is coherently strained to the substrate, with $a$ = 3.85 \AA~and $b$ = 3.86 \AA. The out-of-plane elongation matches the expectation based on the finite thickness of the sample. The unit-cell angles of the \YBCO samples were determined to be 90 degrees by analysing the sets of \{109\} and \{119\} reflections. Therefore, all films are orthorhombic.

The morphology of the samples consists of large two-dimensional islands with a height of 12$\pm$1 \AA, matching the \YBCO $c$-axis lattice parameter. Typically, four different height levels are found within a single terrace of the underlying substrate, indicating that the growth mode of the films is multi-level island growth. Furthermore, some particles with a size of $\sim$10 nm were found on the surface. These particles, which are tentatively attributed to Y$_2$O$_3$ \cite{Catana1993}, are commonly observed on PLD-grown \YBCO and were not further investigated as they are not expected to significantly influence the crystal structure and electrical transport properties.  

\begin{figure}
\includegraphics[width=0.45\textwidth]{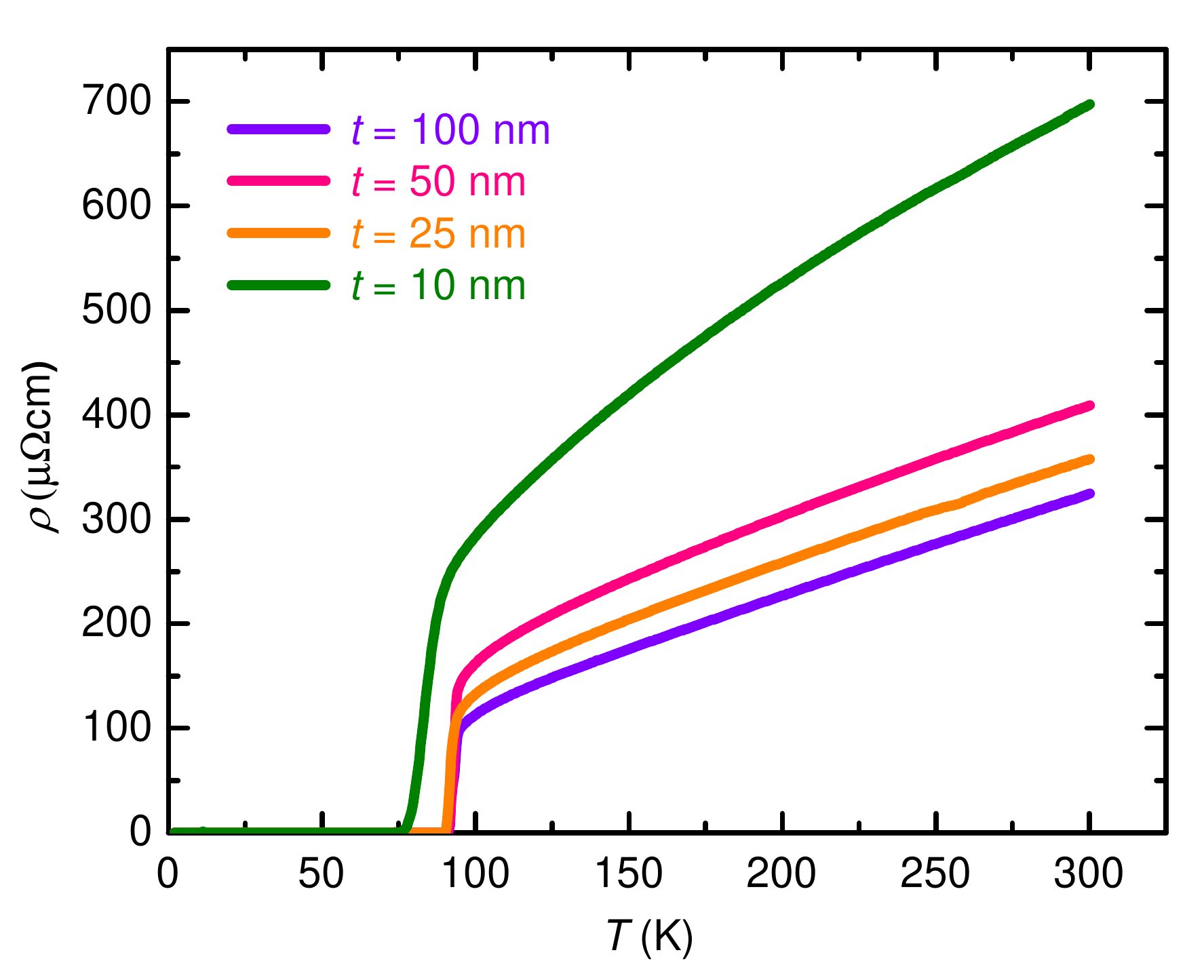}
\caption[]{Temperature dependence of the resistivity of the four \YBCO samples. The three thicker samples have a linear $\rho$($T$) with a $T_\textrm{c}$ of 89 $\pm$ 1 K, whereas the thinnest sample has a sublinear $\rho$($T$) and a reduced $T_\textrm{c}$ of 80 K.}
\label{resis}
\end{figure}

The temperature dependence of the resistivity of all samples is shown in Fig~\ref{resis}. The room-temperature resistivity is $\sim$\SI{350}{\micro\ohm\centi\meter}, the temperature dependence of the resistivity is linear for $T > T_\textrm{c}$, and $T_\textrm{c}$ = 89$\pm$1 K for the films with $t \ge$ 25 nm. These results are in good agreement with the resistivity of optimally doped \YBCO single crystals, albeit with a factor of two higher resistivity \cite{Semba2001, Ando2002}. The thinnest sample, in contrast, has a room temperature resistivity of \SI{700}{\micro\ohm\centi\meter}, a sublinear $\rho$($T$), and a reduced $T_\textrm{c}$ of 80 K. The reduction of $T_\textrm{c}$ in thin \YBCO samples is commonly observed \cite{Triscone1997}.

We start the discussion of the crystal structure of the films and its implications with the observation that the films are untwinned and $c$-axis oriented. Earlier reports mention predominantly $a$-axis oriented films for \YBCO on \NGO at growth temperatures below 750 $^\circ$C and fully $c$-axis oriented films grown above 800 $^\circ$C \cite{Mukaida1993, Jeschke1995, Ece1995}. Moreover, \YBCO thin films on perovskite oxides are generally twinned with up to four different structural variants \cite{Steinborn1994, Schweitzer1996}. The amount of variants can be reduced to two by anisotropic strain using the \NGO (001) crystal orientation \cite{Villard1996} or to one by choosing a suitable miscut of the substrate \cite{Dekkers2003}. Earlier experiments with \NGO (110) resulted in four structural variants \cite{Steinborn1994, Schweitzer1996}. The in-plane coherence of a thin film depends on the initial growth and is limited by anti-phase boundaries \cite{Dekkers2003, Rijnders2004}. The absence of twinning in our films and the $c$-axis orientation is therefore probably explained by an enhanced surface diffusivity due to the chemical substrate preparation resulting in a singly terminated surface. Control experiments using substrates with mixed termination yielded samples with significantly reduced structural quality. However, the general advances in thin-film growth techniques made in the last decades, such as the minimization of impurities and homogeneous ablation of the target, are also expected to be beneficial to the quality of the films.  

The alignment of \YBCO with respect to the optical surface of the substrate is commonly observed \cite{Maurice1998, Dekkers2003}. The large $c$-axis lattice parameter of \YBCO does not match well with the small step-height of the \NGO terraces. The tilt of the \YBCO crystal structure prevents the necessity of anti-phase boundaries occuring at the stepedges, thereby reducing the energy of the film. The tilt, however, also implies a larger substrate-film interface energy as the chemical bonds have to adjust to large local strain fields. The balance of these energies explains why the tilt occurs for the thicker films, but not for the thinner films. 

These \YBCO thin films have an unusual strain state. On average, \NGO (110) is close to being lattice matched and therefore no biaxial strain is present. However, a significant uniaxial strain component is present that compresses the $b$-axis and extends the $a$-axis, thereby reducing the orthorhombicity of the \YBCO from 0.018 to 0.002. This strain state cannot directly be compared to that of \YBCO single crystal subject to uniaxial stress \cite{Welp1992}, because in our case the stress is applied when the \YBCO is cooled through the tetragonal to orthorhombic phase transition at high temperature. During the cooling process, the films remain clamped to the substrates. The orthorhombic-tetragonal phase transition is related to the formation of the CuO chains in the crystal structure. Since the strain state is observed for films with a large thickness, it is very likely that the structure is stabilized by a significant reduction of CuO chain formation. A strain-induced change in oxygen stoichiometry would result in a change in $T_\textrm{c}$\cite{Arpaia2018}, contrary to the observations. Therefore, the oxygen stoichiometry of the films has to be similar to that of optimally doped single crystals and the lack of CuO chains implies most of the oxygen ions in the chain layer are randomly distributed between bonds in the $a$ and $b$ lattice directions. The behavior of these \YBCO films thus implies that a large amount of structural disorder in the chain layers of \YBCO is not detrimental to superconductivity.

In conclusion, we have shown that \YBCO grows coherently strained on \NGO (110) with thicknesses up to 100 nm. The critical temperature is 89$\pm$1 K for $t\ge$ 25 nm. For $t\ge$ 50 nm, the $c$-axis lattice parameter tilts in the direction of the normal to the optical surface of the substrate. These results show that \YBCO grown on \NGO (110) is a promising starting point for systematically studying the superconductivity of the high-$T_\textrm{c}$ cuprates by materials modification using advanced epitaxy. The combination of high $T_\textrm{c}$ and reduced orthorhombicity implies a large amount of structural disorder in the chain layers of \YBCO is not detrimental to superconductivity. Finally, coherently strained \YBCO thin films are expected to be beneficial to superconducting electronic devices, such as grain-boundary Josephson devices \cite{Mannhart1988, Quincey1994}.

We thank Achim G\"uth and Marion Hagel for technical assistance.











\bibliographystyle{aipnum4-1}

\begin{thebibliography}{32}%
\makeatletter
\providecommand \@ifxundefined [1]{%
 \@ifx{#1\undefined}
}%
\providecommand \@ifnum [1]{%
 \ifnum #1\expandafter \@firstoftwo
 \else \expandafter \@secondoftwo
 \fi
}%
\providecommand \@ifx [1]{%
 \ifx #1\expandafter \@firstoftwo
 \else \expandafter \@secondoftwo
 \fi
}%
\providecommand \natexlab [1]{#1}%
\providecommand \enquote  [1]{``#1''}%
\providecommand \bibnamefont  [1]{#1}%
\providecommand \bibfnamefont [1]{#1}%
\providecommand \citenamefont [1]{#1}%
\providecommand \href@noop [0]{\@secondoftwo}%
\providecommand \href [0]{\begingroup \@sanitize@url \@href}%
\providecommand \@href[1]{\@@startlink{#1}\@@href}%
\providecommand \@@href[1]{\endgroup#1\@@endlink}%
\providecommand \@sanitize@url [0]{\catcode `\\12\catcode `\$12\catcode
  `\&12\catcode `\#12\catcode `\^12\catcode `\_12\catcode `\%12\relax}%
\providecommand \@@startlink[1]{}%
\providecommand \@@endlink[0]{}%
\providecommand \url  [0]{\begingroup\@sanitize@url \@url }%
\providecommand \@url [1]{\endgroup\@href {#1}{\urlprefix }}%
\providecommand \urlprefix  [0]{URL }%
\providecommand \Eprint [0]{\href }%
\providecommand \doibase [0]{http://dx.doi.org/}%
\providecommand \selectlanguage [0]{\@gobble}%
\providecommand \bibinfo  [0]{\@secondoftwo}%
\providecommand \bibfield  [0]{\@secondoftwo}%
\providecommand \translation [1]{[#1]}%
\providecommand \BibitemOpen [0]{}%
\providecommand \bibitemStop [0]{}%
\providecommand \bibitemNoStop [0]{.\EOS\space}%
\providecommand \EOS [0]{\spacefactor3000\relax}%
\providecommand \BibitemShut  [1]{\csname bibitem#1\endcsname}%
\let\auto@bib@innerbib\@empty
\bibitem [{\citenamefont {Melhem}(2012)}]{Melhem2012}%
  \BibitemOpen
  \bibinfo {editor} {\bibfnamefont {Z.}~\bibnamefont {Melhem}},\ ed.,\
  \href@noop {} {\emph {\bibinfo {title} {{High Temperature Superconductors
  (HTS) for Energy Applications}}}}\ (\bibinfo  {publisher} {Woodhead
  publishing},\ \bibinfo {year} {2012})\BibitemShut {NoStop}%
\bibitem [{\citenamefont {Khare}(2003)}]{Khare2003}%
  \BibitemOpen
  \bibinfo {editor} {\bibfnamefont {N.}~\bibnamefont {Khare}},\ ed.,\
  \href@noop {} {\emph {\bibinfo {title} {{Handbook of High-Temperature
  Superconducting Electronics}}}}\ (\bibinfo  {publisher} {CRC Press},\
  \bibinfo {year} {2003})\BibitemShut {NoStop}%
\bibitem [{\citenamefont {Tsuei}\ \emph {et~al.}(1994)\citenamefont {Tsuei},
  \citenamefont {Kirtley}, \citenamefont {Chi}, \citenamefont {Yu-Jahnes},
  \citenamefont {Gupta}, \citenamefont {Shaw}, \citenamefont {Sun},\ and\
  \citenamefont {Ketchen}}]{Tsuei1994}%
  \BibitemOpen
  \bibfield  {author} {\bibinfo {author} {\bibfnamefont {C.~C.}\ \bibnamefont
  {Tsuei}}, \bibinfo {author} {\bibfnamefont {J.~R.}\ \bibnamefont {Kirtley}},
  \bibinfo {author} {\bibfnamefont {C.~C.}\ \bibnamefont {Chi}}, \bibinfo
  {author} {\bibfnamefont {L.~S.}\ \bibnamefont {Yu-Jahnes}}, \bibinfo {author}
  {\bibfnamefont {A.}~\bibnamefont {Gupta}}, \bibinfo {author} {\bibfnamefont
  {T.}~\bibnamefont {Shaw}}, \bibinfo {author} {\bibfnamefont {J.~Z.}\
  \bibnamefont {Sun}}, \ and\ \bibinfo {author} {\bibfnamefont {M.~B.}\
  \bibnamefont {Ketchen}},\ }\href {\doibase 10.1103/PhysRevLett.73.593}
  {\bibfield  {journal} {\bibinfo  {journal} {PHYSICAL REVIEW LETTERS}\
  }\textbf {\bibinfo {volume} {73}},\ \bibinfo {pages} {593} (\bibinfo {year}
  {1994})}\BibitemShut {NoStop}%
\bibitem [{\citenamefont {Smilde}\ \emph {et~al.}(2005)\citenamefont {Smilde},
  \citenamefont {Golubov}, \citenamefont {Ariando}, \citenamefont {Rijnders},
  \citenamefont {Dekkers}, \citenamefont {Harkema}, \citenamefont {Blank},
  \citenamefont {Rogalla},\ and\ \citenamefont {Hilgenkamp}}]{Smilde2005}%
  \BibitemOpen
  \bibfield  {author} {\bibinfo {author} {\bibfnamefont {H.}~\bibnamefont
  {Smilde}}, \bibinfo {author} {\bibfnamefont {A.}~\bibnamefont {Golubov}},
  \bibinfo {author} {\bibnamefont {Ariando}}, \bibinfo {author} {\bibfnamefont
  {G.}~\bibnamefont {Rijnders}}, \bibinfo {author} {\bibfnamefont
  {J.}~\bibnamefont {Dekkers}}, \bibinfo {author} {\bibfnamefont
  {S.}~\bibnamefont {Harkema}}, \bibinfo {author} {\bibfnamefont
  {D.}~\bibnamefont {Blank}}, \bibinfo {author} {\bibfnamefont
  {H.}~\bibnamefont {Rogalla}}, \ and\ \bibinfo {author} {\bibfnamefont
  {H.}~\bibnamefont {Hilgenkamp}},\ }\href@noop {} {\bibfield  {journal}
  {\bibinfo  {journal} {{PHYSICAL REVIEW LETTERS}}\ }\textbf {\bibinfo {volume}
  {{95}}} (\bibinfo {year} {{2005}})}\BibitemShut {NoStop}%
\bibitem [{\citenamefont {Kirtley}\ \emph {et~al.}(2006)\citenamefont
  {Kirtley}, \citenamefont {Tsuei}, \citenamefont {Ariando}, \citenamefont
  {Verwijs}, \citenamefont {Harkema},\ and\ \citenamefont
  {Hilgenkamp}}]{Kirtley2006}%
  \BibitemOpen
  \bibfield  {author} {\bibinfo {author} {\bibfnamefont {J.}~\bibnamefont
  {Kirtley}}, \bibinfo {author} {\bibfnamefont {C.}~\bibnamefont {Tsuei}},
  \bibinfo {author} {\bibfnamefont {A.}~\bibnamefont {Ariando}}, \bibinfo
  {author} {\bibfnamefont {C.}~\bibnamefont {Verwijs}}, \bibinfo {author}
  {\bibfnamefont {S.}~\bibnamefont {Harkema}}, \ and\ \bibinfo {author}
  {\bibfnamefont {H.}~\bibnamefont {Hilgenkamp}},\ }\href {\doibase
  {10.1038/nphys215}} {\bibfield  {journal} {\bibinfo  {journal} {{NATURE
  PHYSICS}}\ }\textbf {\bibinfo {volume} {{2}}},\ \bibinfo {pages} {190}
  (\bibinfo {year} {{2006}})}\BibitemShut {NoStop}%
\bibitem [{\citenamefont {Bozovic}\ \emph {et~al.}(2016)\citenamefont
  {Bozovic}, \citenamefont {He}, \citenamefont {Wu},\ and\ \citenamefont
  {Bollinger}}]{Bozovic2016}%
  \BibitemOpen
  \bibfield  {author} {\bibinfo {author} {\bibfnamefont {I.}~\bibnamefont
  {Bozovic}}, \bibinfo {author} {\bibfnamefont {X.}~\bibnamefont {He}},
  \bibinfo {author} {\bibfnamefont {J.}~\bibnamefont {Wu}}, \ and\ \bibinfo
  {author} {\bibfnamefont {A.~T.}\ \bibnamefont {Bollinger}},\ }\href {\doibase
  {10.1038/nature19061}} {\bibfield  {journal} {\bibinfo  {journal} {{NATURE}}\
  }\textbf {\bibinfo {volume} {{536}}},\ \bibinfo {pages} {309} (\bibinfo
  {year} {{2016}})}\BibitemShut {NoStop}%
\bibitem [{\citenamefont {Bousquet}\ \emph {et~al.}(2008)\citenamefont
  {Bousquet}, \citenamefont {Dawber}, \citenamefont {Stucki}, \citenamefont
  {Lichtensteiger}, \citenamefont {Hermet}, \citenamefont {Gariglio},
  \citenamefont {Triscone},\ and\ \citenamefont {Ghosez}}]{Bousquet2008}%
  \BibitemOpen
  \bibfield  {author} {\bibinfo {author} {\bibfnamefont {E.}~\bibnamefont
  {Bousquet}}, \bibinfo {author} {\bibfnamefont {M.}~\bibnamefont {Dawber}},
  \bibinfo {author} {\bibfnamefont {N.}~\bibnamefont {Stucki}}, \bibinfo
  {author} {\bibfnamefont {C.}~\bibnamefont {Lichtensteiger}}, \bibinfo
  {author} {\bibfnamefont {P.}~\bibnamefont {Hermet}}, \bibinfo {author}
  {\bibfnamefont {S.}~\bibnamefont {Gariglio}}, \bibinfo {author}
  {\bibfnamefont {J.-M.}\ \bibnamefont {Triscone}}, \ and\ \bibinfo {author}
  {\bibfnamefont {P.}~\bibnamefont {Ghosez}},\ }\href {\doibase
  {10.1038/nature06817}} {\bibfield  {journal} {\bibinfo  {journal} {{NATURE}}\
  }\textbf {\bibinfo {volume} {{452}}},\ \bibinfo {pages} {{732}} (\bibinfo
  {year} {{2008}})}\BibitemShut {NoStop}%
\bibitem [{\citenamefont {Hwang}\ \emph {et~al.}(2012)\citenamefont {Hwang},
  \citenamefont {Iwasa}, \citenamefont {Kawasaki}, \citenamefont {Keimer},
  \citenamefont {Nagaosa},\ and\ \citenamefont {Tokura}}]{Hwang2012}%
  \BibitemOpen
  \bibfield  {author} {\bibinfo {author} {\bibfnamefont {H.~Y.}\ \bibnamefont
  {Hwang}}, \bibinfo {author} {\bibfnamefont {Y.}~\bibnamefont {Iwasa}},
  \bibinfo {author} {\bibfnamefont {M.}~\bibnamefont {Kawasaki}}, \bibinfo
  {author} {\bibfnamefont {B.}~\bibnamefont {Keimer}}, \bibinfo {author}
  {\bibfnamefont {N.}~\bibnamefont {Nagaosa}}, \ and\ \bibinfo {author}
  {\bibfnamefont {Y.}~\bibnamefont {Tokura}},\ }\href {\doibase
  {10.1038/NMAT3223}} {\bibfield  {journal} {\bibinfo  {journal} {{NATURE
  MATERIALS}}\ }\textbf {\bibinfo {volume} {{11}}},\ \bibinfo {pages} {103}
  (\bibinfo {year} {{2012}})}\BibitemShut {NoStop}%
\bibitem [{\citenamefont {Boschker}\ \emph {et~al.}(2012)\citenamefont
  {Boschker}, \citenamefont {Verbeeck}, \citenamefont {Egoavil}, \citenamefont
  {Bals}, \citenamefont {van Tendeloo}, \citenamefont {Huijben}, \citenamefont
  {Houwman}, \citenamefont {Koster}, \citenamefont {Blank},\ and\ \citenamefont
  {Rijnders}}]{Boschker2012}%
  \BibitemOpen
  \bibfield  {author} {\bibinfo {author} {\bibfnamefont {H.}~\bibnamefont
  {Boschker}}, \bibinfo {author} {\bibfnamefont {J.}~\bibnamefont {Verbeeck}},
  \bibinfo {author} {\bibfnamefont {R.}~\bibnamefont {Egoavil}}, \bibinfo
  {author} {\bibfnamefont {S.}~\bibnamefont {Bals}}, \bibinfo {author}
  {\bibfnamefont {G.}~\bibnamefont {van Tendeloo}}, \bibinfo {author}
  {\bibfnamefont {M.}~\bibnamefont {Huijben}}, \bibinfo {author} {\bibfnamefont
  {E.~P.}\ \bibnamefont {Houwman}}, \bibinfo {author} {\bibfnamefont
  {G.}~\bibnamefont {Koster}}, \bibinfo {author} {\bibfnamefont {D.~H.~A.}\
  \bibnamefont {Blank}}, \ and\ \bibinfo {author} {\bibfnamefont
  {G.}~\bibnamefont {Rijnders}},\ }\href {\doibase {10.1002/adfm.201102763}}
  {\bibfield  {journal} {\bibinfo  {journal} {ADVANCED FUNCTIONAL MATERIALS}\
  }\textbf {\bibinfo {volume} {{22}}},\ \bibinfo {pages} {2235} (\bibinfo
  {year} {{2012}})}\BibitemShut {NoStop}%
\bibitem [{\citenamefont {Lee}\ \emph {et~al.}(2013)\citenamefont {Lee},
  \citenamefont {Orloff}, \citenamefont {Birol}, \citenamefont {Zhu},
  \citenamefont {Goian}, \citenamefont {Rocas}, \citenamefont {Haislmaier},
  \citenamefont {Vlahos}, \citenamefont {Mundy}, \citenamefont {Kourkoutis},
  \citenamefont {Nie}, \citenamefont {Biegalski}, \citenamefont {Zhang},
  \citenamefont {Bernhagen}, \citenamefont {Benedek}, \citenamefont {Kim},
  \citenamefont {Brock}, \citenamefont {Uecker}, \citenamefont {Xi},
  \citenamefont {Gopalan}, \citenamefont {Nuzhnyy}, \citenamefont {Kamba},
  \citenamefont {Muller}, \citenamefont {Takeuchi}, \citenamefont {Booth},
  \citenamefont {Fennie},\ and\ \citenamefont {Schlom}}]{Lee2013}%
  \BibitemOpen
  \bibfield  {author} {\bibinfo {author} {\bibfnamefont {C.-H.}\ \bibnamefont
  {Lee}}, \bibinfo {author} {\bibfnamefont {N.~D.}\ \bibnamefont {Orloff}},
  \bibinfo {author} {\bibfnamefont {T.}~\bibnamefont {Birol}}, \bibinfo
  {author} {\bibfnamefont {Y.}~\bibnamefont {Zhu}}, \bibinfo {author}
  {\bibfnamefont {V.}~\bibnamefont {Goian}}, \bibinfo {author} {\bibfnamefont
  {E.}~\bibnamefont {Rocas}}, \bibinfo {author} {\bibfnamefont
  {R.}~\bibnamefont {Haislmaier}}, \bibinfo {author} {\bibfnamefont
  {E.}~\bibnamefont {Vlahos}}, \bibinfo {author} {\bibfnamefont {J.~A.}\
  \bibnamefont {Mundy}}, \bibinfo {author} {\bibfnamefont {L.~F.}\ \bibnamefont
  {Kourkoutis}}, \bibinfo {author} {\bibfnamefont {Y.}~\bibnamefont {Nie}},
  \bibinfo {author} {\bibfnamefont {M.~D.}\ \bibnamefont {Biegalski}}, \bibinfo
  {author} {\bibfnamefont {J.}~\bibnamefont {Zhang}}, \bibinfo {author}
  {\bibfnamefont {M.}~\bibnamefont {Bernhagen}}, \bibinfo {author}
  {\bibfnamefont {N.~A.}\ \bibnamefont {Benedek}}, \bibinfo {author}
  {\bibfnamefont {Y.}~\bibnamefont {Kim}}, \bibinfo {author} {\bibfnamefont
  {J.~D.}\ \bibnamefont {Brock}}, \bibinfo {author} {\bibfnamefont
  {R.}~\bibnamefont {Uecker}}, \bibinfo {author} {\bibfnamefont {X.~X.}\
  \bibnamefont {Xi}}, \bibinfo {author} {\bibfnamefont {V.}~\bibnamefont
  {Gopalan}}, \bibinfo {author} {\bibfnamefont {D.}~\bibnamefont {Nuzhnyy}},
  \bibinfo {author} {\bibfnamefont {S.}~\bibnamefont {Kamba}}, \bibinfo
  {author} {\bibfnamefont {D.~A.}\ \bibnamefont {Muller}}, \bibinfo {author}
  {\bibfnamefont {I.}~\bibnamefont {Takeuchi}}, \bibinfo {author}
  {\bibfnamefont {J.~C.}\ \bibnamefont {Booth}}, \bibinfo {author}
  {\bibfnamefont {C.~J.}\ \bibnamefont {Fennie}}, \ and\ \bibinfo {author}
  {\bibfnamefont {D.~G.}\ \bibnamefont {Schlom}},\ }\href {\doibase
  {10.1038/nature12582}} {\bibfield  {journal} {\bibinfo  {journal} {{NATURE}}\
  }\textbf {\bibinfo {volume} {{502}}},\ \bibinfo {pages} {{532}} (\bibinfo
  {year} {{2013}})}\BibitemShut {NoStop}%
\bibitem [{\citenamefont {Ghiringhelli}\ \emph {et~al.}(2012)\citenamefont
  {Ghiringhelli}, \citenamefont {Le~Tacon}, \citenamefont {Minola},
  \citenamefont {Blanco-Canosa}, \citenamefont {Mazzoli}, \citenamefont
  {Brookes}, \citenamefont {De~Luca}, \citenamefont {Frano}, \citenamefont
  {Hawthorn}, \citenamefont {He}, \citenamefont {Loew}, \citenamefont {Sala},
  \citenamefont {Peets}, \citenamefont {Salluzzo}, \citenamefont {Schierle},
  \citenamefont {Sutarto}, \citenamefont {Sawatzky}, \citenamefont {Weschke},
  \citenamefont {Keimer},\ and\ \citenamefont {Braicovich}}]{Ghiringhelli2012}%
  \BibitemOpen
  \bibfield  {author} {\bibinfo {author} {\bibfnamefont {G.}~\bibnamefont
  {Ghiringhelli}}, \bibinfo {author} {\bibfnamefont {M.}~\bibnamefont
  {Le~Tacon}}, \bibinfo {author} {\bibfnamefont {M.}~\bibnamefont {Minola}},
  \bibinfo {author} {\bibfnamefont {S.}~\bibnamefont {Blanco-Canosa}}, \bibinfo
  {author} {\bibfnamefont {C.}~\bibnamefont {Mazzoli}}, \bibinfo {author}
  {\bibfnamefont {N.~B.}\ \bibnamefont {Brookes}}, \bibinfo {author}
  {\bibfnamefont {G.~M.}\ \bibnamefont {De~Luca}}, \bibinfo {author}
  {\bibfnamefont {A.}~\bibnamefont {Frano}}, \bibinfo {author} {\bibfnamefont
  {D.~G.}\ \bibnamefont {Hawthorn}}, \bibinfo {author} {\bibfnamefont
  {F.}~\bibnamefont {He}}, \bibinfo {author} {\bibfnamefont {T.}~\bibnamefont
  {Loew}}, \bibinfo {author} {\bibfnamefont {M.~M.}\ \bibnamefont {Sala}},
  \bibinfo {author} {\bibfnamefont {D.~C.}\ \bibnamefont {Peets}}, \bibinfo
  {author} {\bibfnamefont {M.}~\bibnamefont {Salluzzo}}, \bibinfo {author}
  {\bibfnamefont {E.}~\bibnamefont {Schierle}}, \bibinfo {author}
  {\bibfnamefont {R.}~\bibnamefont {Sutarto}}, \bibinfo {author} {\bibfnamefont
  {G.~A.}\ \bibnamefont {Sawatzky}}, \bibinfo {author} {\bibfnamefont
  {E.}~\bibnamefont {Weschke}}, \bibinfo {author} {\bibfnamefont
  {B.}~\bibnamefont {Keimer}}, \ and\ \bibinfo {author} {\bibfnamefont
  {L.}~\bibnamefont {Braicovich}},\ }\href {\doibase {10.1126/science.1223532}}
  {\bibfield  {journal} {\bibinfo  {journal} {{SCIENCE}}\ }\textbf {\bibinfo
  {volume} {{337}}},\ \bibinfo {pages} {{821}} (\bibinfo {year}
  {{2012}})}\BibitemShut {NoStop}%
\bibitem [{\citenamefont {Mankowsky}\ \emph {et~al.}(2014)\citenamefont
  {Mankowsky}, \citenamefont {Subedi}, \citenamefont {Foerst}, \citenamefont
  {Mariager}, \citenamefont {Chollet}, \citenamefont {Lemke}, \citenamefont
  {Robinson}, \citenamefont {Glownia}, \citenamefont {Minitti}, \citenamefont
  {Frano}, \citenamefont {Fechner}, \citenamefont {Spaldin}, \citenamefont
  {Loew}, \citenamefont {Keimer}, \citenamefont {Georges},\ and\ \citenamefont
  {Cavalleri}}]{Mankowsky2014}%
  \BibitemOpen
  \bibfield  {author} {\bibinfo {author} {\bibfnamefont {R.}~\bibnamefont
  {Mankowsky}}, \bibinfo {author} {\bibfnamefont {A.}~\bibnamefont {Subedi}},
  \bibinfo {author} {\bibfnamefont {M.}~\bibnamefont {Foerst}}, \bibinfo
  {author} {\bibfnamefont {S.~O.}\ \bibnamefont {Mariager}}, \bibinfo {author}
  {\bibfnamefont {M.}~\bibnamefont {Chollet}}, \bibinfo {author} {\bibfnamefont
  {H.~T.}\ \bibnamefont {Lemke}}, \bibinfo {author} {\bibfnamefont {J.~S.}\
  \bibnamefont {Robinson}}, \bibinfo {author} {\bibfnamefont {J.~M.}\
  \bibnamefont {Glownia}}, \bibinfo {author} {\bibfnamefont {M.~P.}\
  \bibnamefont {Minitti}}, \bibinfo {author} {\bibfnamefont {A.}~\bibnamefont
  {Frano}}, \bibinfo {author} {\bibfnamefont {M.}~\bibnamefont {Fechner}},
  \bibinfo {author} {\bibfnamefont {N.~A.}\ \bibnamefont {Spaldin}}, \bibinfo
  {author} {\bibfnamefont {T.}~\bibnamefont {Loew}}, \bibinfo {author}
  {\bibfnamefont {B.}~\bibnamefont {Keimer}}, \bibinfo {author} {\bibfnamefont
  {A.}~\bibnamefont {Georges}}, \ and\ \bibinfo {author} {\bibfnamefont
  {A.}~\bibnamefont {Cavalleri}},\ }\href {\doibase {10.1038/nature13875}}
  {\bibfield  {journal} {\bibinfo  {journal} {{NATURE}}\ }\textbf {\bibinfo
  {volume} {{516}}},\ \bibinfo {pages} {{71}} (\bibinfo {year}
  {{2014}})}\BibitemShut {NoStop}%
\bibitem [{\citenamefont {Keimer}\ \emph {et~al.}(2015)\citenamefont {Keimer},
  \citenamefont {Kivelson}, \citenamefont {Norman}, \citenamefont {Uchida},\
  and\ \citenamefont {Zaanen}}]{Keimer2015c}%
  \BibitemOpen
  \bibfield  {author} {\bibinfo {author} {\bibfnamefont {B.}~\bibnamefont
  {Keimer}}, \bibinfo {author} {\bibfnamefont {S.~A.}\ \bibnamefont
  {Kivelson}}, \bibinfo {author} {\bibfnamefont {M.~R.}\ \bibnamefont
  {Norman}}, \bibinfo {author} {\bibfnamefont {S.}~\bibnamefont {Uchida}}, \
  and\ \bibinfo {author} {\bibfnamefont {J.}~\bibnamefont {Zaanen}},\ }\href
  {\doibase {10.1038/nature14165}} {\bibfield  {journal} {\bibinfo  {journal}
  {{NATURE}}\ }\textbf {\bibinfo {volume} {{518}}},\ \bibinfo {pages} {{179}}
  (\bibinfo {year} {{2015}})}\BibitemShut {NoStop}%
\bibitem [{\citenamefont {Badoux}\ \emph {et~al.}(2016)\citenamefont {Badoux},
  \citenamefont {Tabis}, \citenamefont {Laliberte}, \citenamefont
  {Grissonnanche}, \citenamefont {Vignolle}, \citenamefont {Vignolles},
  \citenamefont {Beard}, \citenamefont {Bonn}, \citenamefont {Hardy},
  \citenamefont {Liang}, \citenamefont {Doiron-Leyraud}, \citenamefont
  {Taillefer},\ and\ \citenamefont {Proust}}]{Badoux2016}%
  \BibitemOpen
  \bibfield  {author} {\bibinfo {author} {\bibfnamefont {S.}~\bibnamefont
  {Badoux}}, \bibinfo {author} {\bibfnamefont {W.}~\bibnamefont {Tabis}},
  \bibinfo {author} {\bibfnamefont {F.}~\bibnamefont {Laliberte}}, \bibinfo
  {author} {\bibfnamefont {G.}~\bibnamefont {Grissonnanche}}, \bibinfo {author}
  {\bibfnamefont {B.}~\bibnamefont {Vignolle}}, \bibinfo {author}
  {\bibfnamefont {D.}~\bibnamefont {Vignolles}}, \bibinfo {author}
  {\bibfnamefont {J.}~\bibnamefont {Beard}}, \bibinfo {author} {\bibfnamefont
  {D.~A.}\ \bibnamefont {Bonn}}, \bibinfo {author} {\bibfnamefont {W.~N.}\
  \bibnamefont {Hardy}}, \bibinfo {author} {\bibfnamefont {R.}~\bibnamefont
  {Liang}}, \bibinfo {author} {\bibfnamefont {N.}~\bibnamefont
  {Doiron-Leyraud}}, \bibinfo {author} {\bibfnamefont {L.}~\bibnamefont
  {Taillefer}}, \ and\ \bibinfo {author} {\bibfnamefont {C.}~\bibnamefont
  {Proust}},\ }\href {\doibase {10.1038/nature16983}} {\bibfield  {journal}
  {\bibinfo  {journal} {{NATURE}}\ }\textbf {\bibinfo {volume} {{531}}},\
  \bibinfo {pages} {{210}} (\bibinfo {year} {{2016}})}\BibitemShut {NoStop}%
\bibitem [{\citenamefont {Koster}\ \emph {et~al.}(1998)\citenamefont {Koster},
  \citenamefont {Kropman}, \citenamefont {Rijnders}, \citenamefont {Blank},\
  and\ \citenamefont {Rogalla}}]{Koster1998}%
  \BibitemOpen
  \bibfield  {author} {\bibinfo {author} {\bibfnamefont {G.}~\bibnamefont
  {Koster}}, \bibinfo {author} {\bibfnamefont {B.~L.}\ \bibnamefont {Kropman}},
  \bibinfo {author} {\bibfnamefont {G.~J. H.~M.}\ \bibnamefont {Rijnders}},
  \bibinfo {author} {\bibfnamefont {D.~H.~a.}\ \bibnamefont {Blank}}, \ and\
  \bibinfo {author} {\bibfnamefont {H.}~\bibnamefont {Rogalla}},\ }\href
  {\doibase 10.1063/1.122630} {\bibfield  {journal} {\bibinfo  {journal}
  {APPLIED PHYSICS LETTERS}\ }\textbf {\bibinfo {volume} {73}},\ \bibinfo
  {pages} {2920} (\bibinfo {year} {1998})}\BibitemShut {NoStop}%
\bibitem [{\citenamefont {Catana}\ \emph {et~al.}(1993)\citenamefont {Catana},
  \citenamefont {Bednorz}, \citenamefont {Gerber}, \citenamefont {Mannhart},\
  and\ \citenamefont {Schlom}}]{Catana1993}%
  \BibitemOpen
  \bibfield  {author} {\bibinfo {author} {\bibfnamefont {A.}~\bibnamefont
  {Catana}}, \bibinfo {author} {\bibfnamefont {J.}~\bibnamefont {Bednorz}},
  \bibinfo {author} {\bibfnamefont {C.}~\bibnamefont {Gerber}}, \bibinfo
  {author} {\bibfnamefont {J.}~\bibnamefont {Mannhart}}, \ and\ \bibinfo
  {author} {\bibfnamefont {D.}~\bibnamefont {Schlom}},\ }\href {\doibase
  {10.1063/1.110002}} {\bibfield  {journal} {\bibinfo  {journal} {{APPLIED
  PHYSICS LETTERS}}\ }\textbf {\bibinfo {volume} {{63}}},\ \bibinfo {pages}
  {{553}} (\bibinfo {year} {{1993}})}\BibitemShut {NoStop}%
\bibitem [{\citenamefont {Semba}\ and\ \citenamefont
  {Matsuda}(2001)}]{Semba2001}%
  \BibitemOpen
  \bibfield  {author} {\bibinfo {author} {\bibfnamefont {K.}~\bibnamefont
  {Semba}}\ and\ \bibinfo {author} {\bibfnamefont {A.}~\bibnamefont
  {Matsuda}},\ }\href {\doibase 10.1103/PhysRevLett.86.496} {\bibfield
  {journal} {\bibinfo  {journal} {PHYSICAL REVIEW LETTERS}\ }\textbf {\bibinfo
  {volume} {86}},\ \bibinfo {pages} {496} (\bibinfo {year} {2001})}\BibitemShut
  {NoStop}%
\bibitem [{\citenamefont {Ando}\ \emph {et~al.}(2002)\citenamefont {Ando},
  \citenamefont {Segawa}, \citenamefont {Komiya},\ and\ \citenamefont
  {Lavrov}}]{Ando2002}%
  \BibitemOpen
  \bibfield  {author} {\bibinfo {author} {\bibfnamefont {Y.}~\bibnamefont
  {Ando}}, \bibinfo {author} {\bibfnamefont {K.}~\bibnamefont {Segawa}},
  \bibinfo {author} {\bibfnamefont {S.}~\bibnamefont {Komiya}}, \ and\ \bibinfo
  {author} {\bibfnamefont {A.~N.}\ \bibnamefont {Lavrov}},\ }\href {\doibase
  10.1103/PhysRevLett.88.137005} {\bibfield  {journal} {\bibinfo  {journal}
  {PHYSICAL REVIEW LETTERS}\ }\textbf {\bibinfo {volume} {88}},\ \bibinfo
  {pages} {137005} (\bibinfo {year} {2002})}\BibitemShut {NoStop}%
\bibitem [{\citenamefont {Triscone}\ and\ \citenamefont
  {Fischer}(1997)}]{Triscone1997}%
  \BibitemOpen
  \bibfield  {author} {\bibinfo {author} {\bibfnamefont {J.}~\bibnamefont
  {Triscone}}\ and\ \bibinfo {author} {\bibfnamefont {O.}~\bibnamefont
  {Fischer}},\ }\href {\doibase {10.1088/0034-4885/60/12/004}} {\bibfield
  {journal} {\bibinfo  {journal} {{REPORTS ON PROGRESS IN PHYSICS}}\ }\textbf
  {\bibinfo {volume} {{60}}},\ \bibinfo {pages} {{1673}} (\bibinfo {year}
  {{1997}})}\BibitemShut {NoStop}%
\bibitem [{\citenamefont {Mukaida}\ and\ \citenamefont
  {Miyazawa}(1993)}]{Mukaida1993}%
  \BibitemOpen
  \bibfield  {author} {\bibinfo {author} {\bibfnamefont {M.}~\bibnamefont
  {Mukaida}}\ and\ \bibinfo {author} {\bibfnamefont {S.}~\bibnamefont
  {Miyazawa}},\ }\href {\doibase 10.1063/1.354923} {\bibfield  {journal}
  {\bibinfo  {journal} {JOURNAL OF APPLIED PHYSICS}\ }\textbf {\bibinfo
  {volume} {74}},\ \bibinfo {pages} {1209} (\bibinfo {year}
  {1993})}\BibitemShut {NoStop}%
\bibitem [{\citenamefont {Jeschke}\ \emph {et~al.}(1995)\citenamefont
  {Jeschke}, \citenamefont {Schneider}, \citenamefont {Ulmer},\ and\
  \citenamefont {Linker}}]{Jeschke1995}%
  \BibitemOpen
  \bibfield  {author} {\bibinfo {author} {\bibfnamefont {U.}~\bibnamefont
  {Jeschke}}, \bibinfo {author} {\bibfnamefont {R.}~\bibnamefont {Schneider}},
  \bibinfo {author} {\bibfnamefont {G.}~\bibnamefont {Ulmer}}, \ and\ \bibinfo
  {author} {\bibfnamefont {G.}~\bibnamefont {Linker}},\ }\href {\doibase
  https://doi.org/10.1016/0921-4534(95)00019-4} {\bibfield  {journal} {\bibinfo
   {journal} {PHYSICA C}\ }\textbf {\bibinfo {volume} {243}},\ \bibinfo {pages}
  {243} (\bibinfo {year} {1995})}\BibitemShut {NoStop}%
\bibitem [{\citenamefont {Ece}\ \emph {et~al.}(1995)\citenamefont {Ece},
  \citenamefont {Gonzalez}, \citenamefont {Habermeier},\ and\ \citenamefont
  {Oral}}]{Ece1995}%
  \BibitemOpen
  \bibfield  {author} {\bibinfo {author} {\bibfnamefont {M.}~\bibnamefont
  {Ece}}, \bibinfo {author} {\bibfnamefont {E.~G.}\ \bibnamefont {Gonzalez}},
  \bibinfo {author} {\bibfnamefont {H.}~\bibnamefont {Habermeier}}, \ and\
  \bibinfo {author} {\bibfnamefont {B.}~\bibnamefont {Oral}},\ }\href@noop {}
  {\bibfield  {journal} {\bibinfo  {journal} {JOURNAL OF APPLIED PHYSICS}\
  }\textbf {\bibinfo {volume} {77}},\ \bibinfo {pages} {1646} (\bibinfo {year}
  {1995})}\BibitemShut {NoStop}%
\bibitem [{\citenamefont {Steinborn}\ \emph {et~al.}(1994)\citenamefont
  {Steinborn}, \citenamefont {Miehe}, \citenamefont {Wiesner}, \citenamefont
  {Brecht}, \citenamefont {Fuess}, \citenamefont {Wirth}, \citenamefont
  {Schulte}, \citenamefont {Speckmann}, \citenamefont {Adrian}, \citenamefont
  {Maul}, \citenamefont {Petersen}, \citenamefont {Blau},\ and\ \citenamefont
  {McConnel}}]{Steinborn1994}%
  \BibitemOpen
  \bibfield  {author} {\bibinfo {author} {\bibfnamefont {T.}~\bibnamefont
  {Steinborn}}, \bibinfo {author} {\bibfnamefont {G.}~\bibnamefont {Miehe}},
  \bibinfo {author} {\bibfnamefont {J.}~\bibnamefont {Wiesner}}, \bibinfo
  {author} {\bibfnamefont {E.}~\bibnamefont {Brecht}}, \bibinfo {author}
  {\bibfnamefont {H.}~\bibnamefont {Fuess}}, \bibinfo {author} {\bibfnamefont
  {G.}~\bibnamefont {Wirth}}, \bibinfo {author} {\bibfnamefont
  {B.}~\bibnamefont {Schulte}}, \bibinfo {author} {\bibfnamefont
  {M.}~\bibnamefont {Speckmann}}, \bibinfo {author} {\bibfnamefont
  {H.}~\bibnamefont {Adrian}}, \bibinfo {author} {\bibfnamefont
  {M.}~\bibnamefont {Maul}}, \bibinfo {author} {\bibfnamefont {K.}~\bibnamefont
  {Petersen}}, \bibinfo {author} {\bibfnamefont {W.}~\bibnamefont {Blau}}, \
  and\ \bibinfo {author} {\bibfnamefont {M.}~\bibnamefont {McConnel}},\ }\href
  {\doibase {10.1016/0921-4534(94)90906-7}} {\bibfield  {journal} {\bibinfo
  {journal} {{PHYSICA C}}\ }\textbf {\bibinfo {volume} {{220}}},\ \bibinfo
  {pages} {{219}} (\bibinfo {year} {{1994}})}\BibitemShut {NoStop}%
\bibitem [{\citenamefont {Schweitzer}\ \emph {et~al.}(1996)\citenamefont
  {Schweitzer}, \citenamefont {Bollmeier}, \citenamefont {Stritzker},\ and\
  \citenamefont {Rauschenbach}}]{Schweitzer1996}%
  \BibitemOpen
  \bibfield  {author} {\bibinfo {author} {\bibfnamefont {D.}~\bibnamefont
  {Schweitzer}}, \bibinfo {author} {\bibfnamefont {T.}~\bibnamefont
  {Bollmeier}}, \bibinfo {author} {\bibfnamefont {B.}~\bibnamefont
  {Stritzker}}, \ and\ \bibinfo {author} {\bibfnamefont {B.}~\bibnamefont
  {Rauschenbach}},\ }\href {\doibase {10.1016/0040-6090(95)08210-7}} {\bibfield
   {journal} {\bibinfo  {journal} {{THIN SOLID FILMS}}\ }\textbf {\bibinfo
  {volume} {{280}}},\ \bibinfo {pages} {{147}} (\bibinfo {year}
  {{1996}})}\BibitemShut {NoStop}%
\bibitem [{\citenamefont {Villard}\ \emph {et~al.}(1996)\citenamefont
  {Villard}, \citenamefont {Koren}, \citenamefont {Cohen}, \citenamefont
  {Polturak}, \citenamefont {Thrane},\ and\ \citenamefont
  {Chateignier}}]{Villard1996}%
  \BibitemOpen
  \bibfield  {author} {\bibinfo {author} {\bibfnamefont {C.}~\bibnamefont
  {Villard}}, \bibinfo {author} {\bibfnamefont {G.}~\bibnamefont {Koren}},
  \bibinfo {author} {\bibfnamefont {D.}~\bibnamefont {Cohen}}, \bibinfo
  {author} {\bibfnamefont {E.}~\bibnamefont {Polturak}}, \bibinfo {author}
  {\bibfnamefont {B.}~\bibnamefont {Thrane}}, \ and\ \bibinfo {author}
  {\bibfnamefont {D.}~\bibnamefont {Chateignier}},\ }\href {\doibase
  10.1103/PhysRevLett.77.3913} {\bibfield  {journal} {\bibinfo  {journal}
  {PHYSICAL REVIEW LETTERS}\ }\textbf {\bibinfo {volume} {77}},\ \bibinfo
  {pages} {3913} (\bibinfo {year} {1996})}\BibitemShut {NoStop}%
\bibitem [{\citenamefont {Dekkers}\ \emph {et~al.}(2003)\citenamefont
  {Dekkers}, \citenamefont {Rijnders}, \citenamefont {Harkema}, \citenamefont
  {Smilde}, \citenamefont {Hilgenkamp}, \citenamefont {Rogalla},\ and\
  \citenamefont {Blank}}]{Dekkers2003}%
  \BibitemOpen
  \bibfield  {author} {\bibinfo {author} {\bibfnamefont {J.}~\bibnamefont
  {Dekkers}}, \bibinfo {author} {\bibfnamefont {G.}~\bibnamefont {Rijnders}},
  \bibinfo {author} {\bibfnamefont {S.}~\bibnamefont {Harkema}}, \bibinfo
  {author} {\bibfnamefont {H.}~\bibnamefont {Smilde}}, \bibinfo {author}
  {\bibfnamefont {H.}~\bibnamefont {Hilgenkamp}}, \bibinfo {author}
  {\bibfnamefont {H.}~\bibnamefont {Rogalla}}, \ and\ \bibinfo {author}
  {\bibfnamefont {D.}~\bibnamefont {Blank}},\ }\href {\doibase
  {10.1063/1.1633010}} {\bibfield  {journal} {\bibinfo  {journal} {{APPLIED
  PHYSICS LETTERS}}\ }\textbf {\bibinfo {volume} {{83}}},\ \bibinfo {pages}
  {5199} (\bibinfo {year} {{2003}})}\BibitemShut {NoStop}%
\bibitem [{\citenamefont {Rijnders}\ \emph {et~al.}(2004)\citenamefont
  {Rijnders}, \citenamefont {Curras}, \citenamefont {Huijben}, \citenamefont
  {Blank},\ and\ \citenamefont {Rogalla}}]{Rijnders2004}%
  \BibitemOpen
  \bibfield  {author} {\bibinfo {author} {\bibfnamefont {G.}~\bibnamefont
  {Rijnders}}, \bibinfo {author} {\bibfnamefont {S.}~\bibnamefont {Curras}},
  \bibinfo {author} {\bibfnamefont {M.}~\bibnamefont {Huijben}}, \bibinfo
  {author} {\bibfnamefont {D.}~\bibnamefont {Blank}}, \ and\ \bibinfo {author}
  {\bibfnamefont {H.}~\bibnamefont {Rogalla}},\ }\href {\doibase
  {10.1063/1.1646463}} {\bibfield  {journal} {\bibinfo  {journal} {{APPLIED
  PHYSICS LETTERS}}\ }\textbf {\bibinfo {volume} {{84}}},\ \bibinfo {pages}
  {1150} (\bibinfo {year} {{2004}})}\BibitemShut {NoStop}%
\bibitem [{\citenamefont {Maurice}\ \emph {et~al.}(1998)\citenamefont
  {Maurice}, \citenamefont {Durand}, \citenamefont {Drouet},\ and\
  \citenamefont {Contour}}]{Maurice1998}%
  \BibitemOpen
  \bibfield  {author} {\bibinfo {author} {\bibfnamefont {J.-L.}\ \bibnamefont
  {Maurice}}, \bibinfo {author} {\bibfnamefont {O.}~\bibnamefont {Durand}},
  \bibinfo {author} {\bibfnamefont {M.}~\bibnamefont {Drouet}}, \ and\ \bibinfo
  {author} {\bibfnamefont {J.-P.}\ \bibnamefont {Contour}},\ }\href {\doibase
  https://doi.org/10.1016/S0040-6090(97)01124-3} {\bibfield  {journal}
  {\bibinfo  {journal} {THIN SOLID FILMS}\ }\textbf {\bibinfo {volume} {319}},\
  \bibinfo {pages} {211} (\bibinfo {year} {1998})}\BibitemShut {NoStop}%
\bibitem [{\citenamefont {Welp}\ \emph {et~al.}(1992)\citenamefont {Welp},
  \citenamefont {Grimsditch}, \citenamefont {Fleshler}, \citenamefont
  {Nessler}, \citenamefont {Downey}, \citenamefont {Crabtree},\ and\
  \citenamefont {Guimpel}}]{Welp1992}%
  \BibitemOpen
  \bibfield  {author} {\bibinfo {author} {\bibfnamefont {U.}~\bibnamefont
  {Welp}}, \bibinfo {author} {\bibfnamefont {M.}~\bibnamefont {Grimsditch}},
  \bibinfo {author} {\bibfnamefont {S.}~\bibnamefont {Fleshler}}, \bibinfo
  {author} {\bibfnamefont {W.}~\bibnamefont {Nessler}}, \bibinfo {author}
  {\bibfnamefont {J.}~\bibnamefont {Downey}}, \bibinfo {author} {\bibfnamefont
  {G.~W.}\ \bibnamefont {Crabtree}}, \ and\ \bibinfo {author} {\bibfnamefont
  {J.}~\bibnamefont {Guimpel}},\ }\href {\doibase 10.1103/PhysRevLett.69.2130}
  {\bibfield  {journal} {\bibinfo  {journal} {PHYSICAL REVIEW LETTERS}\
  }\textbf {\bibinfo {volume} {69}},\ \bibinfo {pages} {2130} (\bibinfo {year}
  {1992})}\BibitemShut {NoStop}%
\bibitem [{\citenamefont {Arpaia}\ \emph {et~al.}(2018)\citenamefont {Arpaia},
  \citenamefont {Andersson}, \citenamefont {Trabaldo}, \citenamefont {Bauch},\
  and\ \citenamefont {Lombardi}}]{Arpaia2018}%
  \BibitemOpen
  \bibfield  {author} {\bibinfo {author} {\bibfnamefont {R.}~\bibnamefont
  {Arpaia}}, \bibinfo {author} {\bibfnamefont {E.}~\bibnamefont {Andersson}},
  \bibinfo {author} {\bibfnamefont {E.}~\bibnamefont {Trabaldo}}, \bibinfo
  {author} {\bibfnamefont {T.}~\bibnamefont {Bauch}}, \ and\ \bibinfo {author}
  {\bibfnamefont {F.}~\bibnamefont {Lombardi}},\ }\href@noop {} {\bibfield
  {journal} {\bibinfo  {journal} {{PHYSICAL REVIEW MATERIALS}}\ }\textbf
  {\bibinfo {volume} {{2}}} (\bibinfo {year} {{2018}})}\BibitemShut {NoStop}%
\bibitem [{\citenamefont {Mannhart}\ \emph {et~al.}(1988)\citenamefont
  {Mannhart}, \citenamefont {Chaudhari}, \citenamefont {Dimos}, \citenamefont
  {Tsuei},\ and\ \citenamefont {McGuire}}]{Mannhart1988}%
  \BibitemOpen
  \bibfield  {author} {\bibinfo {author} {\bibfnamefont {J.}~\bibnamefont
  {Mannhart}}, \bibinfo {author} {\bibfnamefont {P.}~\bibnamefont {Chaudhari}},
  \bibinfo {author} {\bibfnamefont {D.}~\bibnamefont {Dimos}}, \bibinfo
  {author} {\bibfnamefont {C.~C.}\ \bibnamefont {Tsuei}}, \ and\ \bibinfo
  {author} {\bibfnamefont {T.~R.}\ \bibnamefont {McGuire}},\ }\href {\doibase
  10.1103/PhysRevLett.61.2476} {\bibfield  {journal} {\bibinfo  {journal}
  {PHYSICAL REVIEW LETTERS}\ }\textbf {\bibinfo {volume} {61}},\ \bibinfo
  {pages} {2476} (\bibinfo {year} {1988})}\BibitemShut {NoStop}%
\bibitem [{\citenamefont {Quincey}(1994)}]{Quincey1994}%
  \BibitemOpen
  \bibfield  {author} {\bibinfo {author} {\bibfnamefont {P.~G.}\ \bibnamefont
  {Quincey}},\ }\href {\doibase 10.1063/1.111091} {\bibfield  {journal}
  {\bibinfo  {journal} {APPLIED PHYSICS LETTERS}\ }\textbf {\bibinfo {volume}
  {64}},\ \bibinfo {pages} {517} (\bibinfo {year} {1994})}\BibitemShut
  {NoStop}%
\end{thebibliography}

\end{document}